# Glue-Assisted Grinding Exfoliation of Large-Size 2D Materials for Insulating Thermal Conduction and Large-Current-Density Hydrogen Evolution

Liusi Yang[1], Dashuai Wang[1], Minsu Liu[1], Heming Liu[1], Junyang Tan[1], Heyuan Zhou[1], Zhongyue Wang[1], Qiangmin Yu[1], Jingyun Wang[1], Junhao Lin[3], Xiaolong Zou[1], Ling Qiu[1], Hui-Ming Cheng[1,2], Bilu Liu[1]*

[1]Shenzhen Geim Graphene Center, Tsinghua-Berkeley Shenzhen Institute & Institute of Materials Research, Tsinghua Shenzhen International Graduate School, Tsinghua University, Shenzhen 518055, P. R. China.
[2] Laboratory for Materials Sciences, Institute of Metal Research, Chinese Academy of Sciences, Shenyang 110016, P. R. China
[3] Department of Physics, Southern University of Science and Technology, Shenzhen 518055, P. R. China
*E-mail: bilu.liu@sz.tsinghua.edu.cn (B.L.)

**Abstract**

Two-dimensional (2D) materials have many promising applications, but their scalable production remains challenging. Herein, we develop a glue-assisted grinding exfoliation (GAGE) method in which the adhesive polymer acts as a glue to massively produce 2D materials with large lateral sizes, high quality, and high yield. Density functional theory simulation shows that the exfoliation mechanism involves the competition between the binding energy of selected polymers and the 2D materials which is larger than the exfoliation energy of the layered materials. Taking h-BN as an example, the GAGE produces 2D h-BN with an average lateral size of 2.18 μm and thickness of 3.91 nm. The method is also extended to produce various other 2D materials, including graphene, $MoS_2$, $Bi_2O_2Se$, vermiculite, and montmorillonite. Two representative applications of thus-produced 2D materials have been demonstrated, including h-BN/polymer composites for insulating thermal conduction and $MoS_2$ electrocatalysts for large-current-density hydrogen evolution, indicating the great potential of massively produced 2D materials.

**Introduction**

Two-dimensional (2D) materials with intriguing physical and chemical properties are expected to have a variety of applications in thermal management, energy conversion, flexible devices, functional composites, *etc.* [1-3] For these applications, the prerequisite is the efficient and scalable production of 2D materials with high quality and low cost. Many efforts to achieve this goal have been reported, and "top-down" exfoliation techniques have shown great promise to realize the low-cost production of 2D materials with high yield by overcoming the interlayer interactions in layered materials.[4-8] For example, ultrasonication and high-shear mixing can produce 2D nanosheets from their bulk counterparts, but the yields are relatively low.[9-12] Ball milling has a higher yield, but the nanosheets are usually small and contain defects due to the high energy impact.[13,14] Electrochemical exfoliation can prepare monolayer 2D nanosheets with a high yield, while it is difficult to scale up and mainly work for conductive materials.[15,16] The intermediate assisted grinding exfoliation (iMAGE) technique is efficient in producing 2D materials with high quality, but obtaining larger and thinner 2D sheets is hampered by the weak interaction between the intermediate (e.g., silicon carbide particles) and the layered materials, and the particles need to be removed before the layered materials can be used.[17] The development of universal high-yield exfoliation methods, which can produce ultrathin 2D nanosheets with a large lateral size and high quality, remains challenging.

Interestingly, we note that the micromechanical exfoliation method which was used in the discovery of graphene produces monolayer 2D sheets with the highest quality but cannot be scaled up.[18] In this method, Scotch® tape adheres to the surface of layered materials and exfoliates them with an external force because the interaction between the Scotch® tape and layered materials is stronger than the interlayer interaction of the layered materials. In essence, it is the adhesive polymer (e.g., rubber, acrylic) on the tape that plays the key role as a bonding agent. Therefore, using an adhesive polymer in a more suitable medium has huge potential to exfoliate layered materials with higher efficiency and remain their highest quality. Besides, the adhesive polymers have been used in a variety of applications related to 2D materials, such as binders for Li-ion batteries, mechanical reinforcement, dispersing agents for sensors, and hydrogel for drug delivery.[19-24] Hence, using adhesive polymers as exfoliation agents may realize "one-step" exfoliation and dispersion of high-quality 2D materials, as well as direct preparation in functional composites without removing the polymer for subsequent applications.

Inspired by the above facts, here we report an effective glue-assisted grinding exfoliation (GAGE) approach that uses an adhesive polymer solution as the glue to massively produce 2D materials. Taking hexagonal boron nitride (h-BN) as an example, we achieved the controllable preparation of ultrathin boron nitride nanosheets (BNNSs) with large lateral sizes, high quality, and high yield. The exfoliation occurs because the binding energy between the adhesive polymer and h-BN is larger than the exfoliation energy of the layered materials, which is confirmed by density functional theory (DFT) simulations. This method can be used to prepare many other 2D materials, including conducting graphene, semiconducting molybdenum disulfide ($MoS_2$) and $Bi_2O_2Se$, and insulating clay materials. The 2D material/polymer composite dispersion produced by the method showed remarkable performance in two scalable applications. First, a BNNS-based composite film has high thermal conductivity, good mechanical performance, and electrical insulation for flexible thermal management. Second, a 2D $MoS_2$-based electrocatalyst has high activity and good durability for large-current-density hydrogen evolution reaction (HER). The results suggest the great potential of the massively produced 2D materials in applications.

**Results and Discussion**

Inspired by the micromechanical exfoliation method, our idea originates from using an adhesive polymer solution as a "glue" directly to replace the adhesive tape and introducing shear forces by an automatic grinding process to produce 2D materials with large lateral size. h-BN was chosen as the representative material because it is useful in insulating thermal conduction.[25,26] To select the adhesive polymer, we analyzed the adhesion and exfoliation mechanism by DFT simulations and taken two types of adhesive polymer as examples. [27-30] As shown in Figure 1a, polyethyleneimine (PEI) with a long-chain structure adheres to the surface of h-BN by electrostatic interaction. For another, carboxymethyl cellulose (CMC), which is an aromatic structure, has stronger electrostatic interactions with the surface of h-BN sheets by chemical adsorption of Na atoms in branches of CMC with N atoms (Figure 1b). The binding energies between h-BN and the polymers were calculated to be 0.210 J m$^{-2}$ for h-BN/PEI and 0.450 J m$^{-2}$ for h-BN/CMC (Figure 1c). The exfoliation energy of bulk h-BN was also calculated to be 0.195 J m$^{-2}$ by comparing the energy difference between bulk h-BN and its monolayer counterpart (Figure S1). As a result, the binding energies at the h-BN/polymer interface are larger than the exfoliation energy of bulk h-BN, showing the feasibility of exfoliating h-BN with the assistance of polymer, due to the stronger interactions between the selected polymer and h-BN than the interlayer interaction of bulk h-BN.

Guided by the DFT simulations, we have developed a GAGE method for the production of 2D materials (Figure 1d). When the adhesive polymer solution and the layered material are mixed and ground by a mortar grinder, the adhesive polymers contact and interact with the surface layer of the bulk materials, and shear force originated from the movement of the pestle and mortar causes sliding and exfoliation of these layers through overcoming the interlayer forces. In the GAGE method, combined with the shear force from the grinding process, the highly viscous polymer solution with adhesive characteristics would be effective to prepare 2D materials.

By using water-soluble CMC as the adhesive "glue", bulk h-BN powder was exfoliated by the GAGE method. Scanning electron microscope (SEM) images show that the rigid and thick bulk h-BN powders were transformed into flexible and thin BNNSs after the GAGE process (Figure 2a, b). The obvious Tyndall effect of the BNNSs dispersed in water indicates its good dispersibility (Inset of Figure 2b). High-resolution transmission electron microscope (HRTEM) images focused on the edges of the BNNSs show their ultrathin structure with few layers (Figure 2c, S2). Also, the HRTEM image of the basal plane and the hexagonally arranged diffraction spots in the selected area electron diffraction (SAED) pattern suggest the high crystallinity of the BNNSs without noticeable defects (Figure 2d, S3). The morphology of BNNSs was further investigated by atomic force microscopy (AFM) and the corresponding height profile, which confirms the large lateral size and thin thickness of the 2D products (Figure 2e). Statistical TEM and AFM analysis show that the BNNSs have a lateral size distribution of 2.18±0.75 μm, in accordance with the dynamic light scattering (DLS) results, and thickness distribution of 3.91±1.74 nm with a high yield of 32.8% (Figure 2f, g, S4). Compared with other exfoliation methods, including ultrasonication, ball milling, intermediate-assisted grinding exfoliation (iMAGE), *etc.* the GAGE method produced BNNSs with the largest lateral size and aspect ratio (Figure 2h, Table S1).[4,6,11,13,17,31,32]

We also studied the chemical composition and structure of the BNNSs by X-ray photoelectron spectroscopy (XPS) and powder X-ray diffraction (XRD). The characteristic peaks for sodium and oxygen from CMC had disappeared as shown by the XPS spectra, suggesting that the polymers were removed completely after washing and annealing treatment (Figure 2i). Moreover, the boron and nitrogen peaks in the high-resolution XPS spectra show no obvious changes between bulk h-BN powders and BNNSs, indicating that the BNNSs maintained the original structure with no additional functional groups (Figure 2j, k).[6] In the XRD pattern, the slight shift of the (0 0 2) peak from 26.87° for bulk h-BN to 26.76° for the BNNSs corresponds to an increase in the interlayer spacing after exfoliation (Figure 2l).[6] The results confirmed that high-quality BNNSs were produced by the GAGE strategy.

In addition to CMC, other adhesive polymers, such as PEI, sodium alginate, soluble starch, polyacrylamide (PAM), and polyvinyl pyrrolidone (PVP) can also be used as glue media to prepare 2D materials. They are water-soluble with high viscosities and can interact with layered materials.[21,33] Stable BNNS dispersions formed after the GAGE process using the above polymers are shown in Figure S5. In addition to the successful exfoliation of insulating h-BN, we found that this method also could produce other 2D materials from their layered counterparts (Figure S6). Conducting graphene, semiconducting $MoS_2$, $WS_2$, and $Bi_2O_2Se$, and insulating clay materials (mica, vermiculite, and montmorillonite) were prepared as homogeneous 2D sheets dispersions and the UV-vis-NIR absorption spectra present the characteristic absorption peaks of these 2D materials (Figure 3a, b).[17,34,35]

Regarding the exfoliation of other 2D materials, we used graphene and $MoS_2$ as examples. Based on DFT simulations, the binding energy between graphene and PEI (0.276 J m$^{-2}$) is larger than the exfoliation energy of graphite (0.165 J m$^{-2}$), which provides theoretical support for the experiment (Figure S7). Similarly, the binding energy between MoS$_2$ and CMC (0.465 J m$^{-2}$) is larger than the exfoliation energy of MoS$_2$ (0.222 J m$^{-2}$, Figure S8). Concerning the morphology and quality of the exfoliated products, the ultrathin structure and high crystallinity of the graphene sheets are shown by the low contrast in the SEM and TEM images and clear diffraction spots in the SAED pattern (Figure 3c, S9). Statistical analysis reveals the average lateral size of the as-obtained graphene is 2.74 μm with a standard deviation of 0.75 μm (Figure 3d). Raman spectra of the exfoliated graphene have a typical D band at 1347 cm$^{-1}$, a G band at 1580 cm$^{-1}$, and a 2D band at 2718 cm$^{-1}$.[12] A low D/G intensity ratio indicates the graphene sheets are high-quality and maintain the original structure with no oxidation, which is also confirmed by the XPS spectra (Figure 3e, S10).

For the exfoliated MoS$_2$ nanosheets, the average lateral size is ~336 nm according to the DLS results and clear lattice fringes in the HRTEM images suggest a thickness of ~8 nm (Figure S11,12). The Raman spectroscopy shows two characteristic peaks of MoS$_2$, i.e., E$_{2g}$ at 383 cm$^{-1}$ and A$_{1g}$ at 408 cm$^{-1}$ (Figure S13).[15,36] We also studied the quality of the exfoliated MoS$_2$ by high-angle annular dark-field scanning transmission electron microscopy (HAADF-STEM). The separated edges of the MoS$_2$ nanosheets and the well-organized atomic structure strongly indicate the successful exfoliation and high crystallinity of the nanosheets without noticeable defects (Figure 3f, g). Other materials, WS$_2$, Bi$_2$O$_2$Se, mica, vermiculite, and montmorillonite, can also be exfoliated into 2D sheets with lateral sizes from hundreds of nanometers to tens of micrometers, depending on the original size and type of the raw material, type of polymer, and the grinding time (Figure S14). In brief, the above results confirm the universality of the GAGE method for preparing abundant and high-quality 2D nanosheets.

We have performed systematic experiments and summarized the three roles of the polymers in the GAGE method. First, polymers act as glues to interact with the layered materials and provide viscous adhesives. Second, polymers can encapsulate the 2D material sheets, provide effective separation between the sheets and allow enhanced exfoliation. Third, polymers are usually dispersing agents and prevent the 2D material sheets from re-aggregation, thus forming a homogeneous dispersion. Therefore, using a mild grinding process to introduce the shear forces, adhesive polymers are able to produce 2D materials with larger lateral size and thinner thickness.

Owing to the feasibility of the GAGE method for the production of high-quality 2D materials in large quantities, we explored the potential use of the 2D material/polymer composite products in two scalable applications. With the advent of the 5G era, there is an increasing requirement for thermal interface materials to improve the heat-dissipation of high-power density electronic devices. Considering their high thermal conductivity, electrical insulation, and excellent stability, BNNSs are ideal nanofillers to fabricate polymer composites with good processability for thermal management.[25,37] To demonstrate its adaptability for scale-up to an industrial level, we used the "one-step" GAGE method to prepare 10 L BNNS/CMC dispersions and achieved simultaneous exfoliation and dispersion (Figure 4a). Different from the method of adding a small amount of h-BN filler to the polymer matrix, our composite dispersions contain higher contents of the BNNS than polymer, which are usually larger than 60 wt%. A viscous composite dispersion with a high h-BN concentration of ~80 mg mL$^{-1}$ was assembled by doctor-blade casting, and after drying we obtained a ~200 cm$^2$ free-standing BNNS/CMC composite film where the BNNSs were arranged orderly (Figure S15).[38] For example, a composite film with a BNNS loading of 60 wt% could be easily folded into an "airplane" or tied into a knot, demonstrating its good flexibility and robust mechanical property even at a high loading of BNNS. We also studied the mechanical properties of the BNNS/CMC composite films by tensile tests. With increasing the BNNS content from 60 to 90 wt%, the tensile strength initially increased and then decreased and ranged from 12.5 MPa to 17.8 MPa (Figure S16). The Young's modulus showed the same trend with a maximum of 3254±462 MPa at a BNNS loading of 80 wt% (Figure 4b).

For the thermal conductivity of the BNNS/CMC composite film, it depends on the content of BNNS, and the in-plane thermal conductivity reaches 25.92 W m$^{-1}$·K$^{-1}$ at a BNNS loading of 90 wt%, which is more than 20 times that of a pure CMC film (1.29 W m$^{-1}$·K$^{-1}$, Figure 4c). Meanwhile, the thermal diffusion coefficient was 14.73 mm$^2$ s$^{-1}$. The achievement of a high BNNS content indicates good wettability between BNNS and CMC that contributes to the formation of a homogeneous heat transfer network.[39] Moreover, the CMC promoted a high viscosity of the fluid, which is a dominant factor to obtain high shear stress in doctor-blade casting.[40] It is suspected that the high shear stress orients the BNNSs, facilitating the enhancement of thermal conductivity. As summarized in Figure 4d and Table S2, compared with other previously reported BNNS/polymer composite films, the BNNS/CMC composite film produced by the GAGE method and doctor-blading has the highest thermal conductivity.[13,25,37,41-47] This ultrahigh thermal conductivity is attributed to the presence of ultrathin BNNS with large lateral sizes and can be improved by sieving out the unexfoliated h-BN sheets. Figure 4e depicts the frequency-dependent dielectric properties and shows that the BNNS/CMC composite films have a better insulating property with a lower electrical conductivity compared with the pure CMC film. Therefore, the BNNS/CMC composite film exhibits good mechanical flexibility, ultrahigh thermal conductivity, and enhanced electrical insulation.

We also used the "one-step" GAGE method for another scalable application in the energy-related field. $MoS_2$ is a promising electrocatalyst for the HER owing to its high catalytic activity, low cost, and good stability.[48] Compared to bulk $MoS_2$, which is an abundant natural resource, 2D $MoS_2$ nanosheets have more exposed active edge sites and higher catalytic activity,[49] however, their efficient and low-cost production remains challenging. As a result, we chose cheap industrial-grade $MoS_2$ powder from a mineral company as the raw material and used the GAGE method to prepare 2D $MoS_2$ dispersions. Taking advantage of the exfoliation and dispersing agents of ethyl cellulose (EC), after the GAGE process the 2D $MoS_2$/EC dispersion is very stable for 60 days, which would promote the uniform coating of $MoS_2$ catalysts on a substrate (Figure 5a). In contrast, a dispersion containing the original $MoS_2$ powders has obvious precipitates after 10 hours. The zeta potential of the 2D $MoS_2$/EC dispersions (-37.8 mV) confirms its good dispersibility, compared with that of $MoS_2$ powder dispersions (-11.4 mV, Figure S17). For the fabrication of 2D $MoS_2$-based electrocatalysts, the stable 2D $MoS_2$/EC dispersion was dropped into a porous Cu foam, followed by a simple annealing treatment. After the annealing process, 2D $MoS_2$ sheets were still uniformly anchored to the substrate as shown in the SEM images, and the excess polymer residue was removed as confirmed by the thermogravimetric analyzer (TGA) curve of the EC powder (Figure S18, 19).

To meet the demands of industrial applications, we tested the HER performance of the 2D $MoS_2$-based catalyst in alkaline media at large current densities ($j$). In a KOH (1M) electrolyte, the polarization curves show that the overpotential ($\eta$) of the 2D $MoS_2$-based electrocatalyst at 1000 mA

cm$^{-2}$ is 409 mV, which is close to that of the Pt/C catalyst loaded on Cu foam at a loading of 2 mg cm$^{-2}$ (410 mV@1000 mA cm$^{-2}$, Figure 5b, c), as well as the state-of-the-art performance reported for Pt-based catalysts.[50] For practical HER applications, the large-current-density performance of the electrocatalyst is critical, which can be evaluated by analyzing the relationship between current density and $\Delta\eta/\Delta\log|j|$. As the current density increased, the $\Delta\eta/\Delta\log|j|$ ratio of the Pt/C catalyst increased sharply, however, that of the MoS$_2$-based catalyst maintained a small value, suggesting its good performance at different large current densities (Figure 5d). Compared with the Pt-based catalysts and MoS$_2$-based composite catalysts reported in the literature, the HER performance of our cheap and mass-produced 2D MoS$_2$-based catalysts is comparable to their optimal performance at a large current density of 1000 mA cm$^{-2}$ (Figure 5e, Table S3).[48-54] Furthermore, after 10,000 cycles, the polarization curve of the 2D MoS$_2$-based catalyst is similar to the initial one, and the catalysts remained stable after a 24 h current-time test at 100 mA cm$^{-2}$, indicating its good durability for long-term use (Figure 5f). The above results confirmed that the 2D MoS$_2$-based catalysts prepared by the GAGE method had a good HER performance with a small overpotential and good durability at large current densities.

**Conclusion**

In summary, we have developed an effective and universal GAGE method for the scalable production of 2D materials with large lateral size and high quality by using an adhesive polymer solution as the grinding medium. DFT simulations show that the binding energies between the adhesive polymer and the 2D material are larger than the exfoliation energies of the layered materials, providing the basis for successful exfoliation. Furthermore, the 2D material/polymer composite dispersions produced by the GAGE method have shown remarkable performance in two scalable applications, including flexible BNNS/CMC composite films with ultrahigh thermal conductivity for insulating thermal conduction and MoS$_2$-based electrocatalysts with a small overpotential and good durability for the large-current-density HER. The GAGE method has huge potential for the mass-production of high-quality novel 2D materials and the construction of functional polymer composites for many applications, such as actuators, energy storage devices, and electronic devices.

**Experimental Section**

**Materials.** h-BN (Tanyun Junrong Liaoning Chemical Research Institute New Materials Incubator Co., Ltd., D$_{50}$ of 30 μm), graphite (Macklin Biochemical Co. Ltd., 99.95%, 325 mesh), MoS$_2$ (Aladdin Chemical Reagent Co., 99.5%, <2 μm), WS$_2$ (Macklin Biochemical Co. Ltd., 99.9%, 2 μm), Vermiculite (Macklin Biochemical Co. Ltd., 95%), Montmorillonite (Macklin Biochemical Co. Ltd., K-10), industrial-grade MoS$_2$ (Luoyang Shenyu Molybdenum Co., LTD, D$_{50}$ of 16~30 μm), and PEI solution (Sigma-Aldrich, average Mn~60000 by GPC, average Mw~750000 by LS, 50 wt% in H$_2$O), were used as received. The bulk Bi$_2$O$_2$Se crystal was synthesized by the chemical vapor transfer method as previously reported and then ground into powders by hand.[35] Mica (Chuzhou Grea Minerals Co., Ltd., GM-5) was first annealed at 800 in the air for 1 h.[34] Adhesive polymer powders, CMC (Macklin, viscosity 5000-15000 mPa·s, USP grade), sodium alginate (Sigma-Aldrich, Medium viscosity) soluble starch (Sigma-Aldrich, ACS reagent), PAM (Macklin, non-ionic, M~7000000), PVP (Aladdin Chemical Reagent Co., average M~58000, K29-32), were dissolved in deionized water to form homogeneous solutions with high viscosities (2000-30000 mPa·s) using an electric mixer (Shanghai Huxi, RWD 150) for high-speed stirring. Similarly, EC (Macklin Biochemical Co. Ltd.,) was dissolved in ethanol.

**Preparation of 2D materials by the GAGE method.** In a typical GAGE process, taking the exfoliation of h-BN by CMC solution as an example, the bulk h-BN powder (7 g) and the CMC solution (150 g, the content of CMC is 3 g) were added to a mortar grinder and ground for 9 h (Retsch, RM 200). During the grinding process, the mortar was rotated at a speed of 100 rpm, and the h-BN powders were exfoliated into 2D BNNS in an ambient atmosphere. The homogeneous BNNS/CMC dispersion was obtained and used directly for applications. For the other selected 2D materials and polymers, the GAGE process was also performed in an ambient atmosphere with controllable parameters, including grinding time, and the weight ratio of bulk materials and polymers. For the morphology characterization of the 2D materials, most of the polymer was removed by multiple cycles of centrifugation at a speed of 10,000 rpm (10619$g$) and washing treatments in deionized water, followed by the annealing treatment that completely removed any remaining residue. The yield of 2D materials was calculated using the mass of the exfoliated material from the supernatant after centrifugation at a speed of 3000 rpm (955$g$) for 15 min.

**Theoretical calculations of adhesion energies.** DFT calculations were performed with the Vienna ab initio Simulation Package (VASP).[27] The generalized gradient approximation proposed by Perdew, Burke, and Ernzerh (PBE) and the projector augmented wave (PAW) method were adopted to describe exchange-correlation and ion-electron interactions.[28,29] A plane-wave cutoff energy was set to 400 eV. Computationally cost-effective Grimme's D3 method was used to determine weak van der Waals interactions.[30] For the energy calculations of graphene, h-BN monolayer with $12 \times 7\sqrt{3}$ supercell, and MoS$_2$ monolayer with $10 \times 6\sqrt{3}$ supercell, the Brillouin-zone integration was performed using only a 1×1×1 Gamma-centered $k$-point sampling. To avoid spurious interactions periodic images, the vacuum separation between two neighboring monolayers was set to 20 Å. Ionic and electronic relaxations were performed until a convergence criterion of 0.1 eV/Å per ion and $10^{-4}$ eV per electronic step were achieved, respectively.

**Assembly of the BNNS/CMC composite films.** A continuous BNNS/CMC film was fabricated by a simple doctor-blade casting technique. Typically, a lab-scale doctor blade coater (KJ-MTI, MSK-AFA-IIID) was used to spread 10 mL of the homogeneous BNNS/CMC dispersion over a silicone-coated PET release liner (release force: 5-10 N). The casting speed of the coater was set to 50 mm s$^{-1}$, and the gap of the doctor blade was controlled in the range of 100-1000 μm to obtain a wet film. The wet film was dried in a vacuum oven at 60ºC and 100 Pa for 12 hours. Afterward, the dried film was then hot-pressed under a pressure of 15 MPa and a temperature of 100ºC for 10 min. Finally, the PET release liner was removed, and a free-standing and flexible BNNS/CMC film with a thickness in the range of 47-63 μm was obtained.

**Fabrication of 2D MoS$_2$-based catalysts and electrochemical measurements.** First, the industrial-grade MoS$_2$ powder (2 g) was exfoliated and dispersed by EC solutions using the GAGE process. Second, the 2D MoS$_2$/EC dispersion was dropped into a Cu foam (1×1.5 cm$^2$) with a controlled loading of 10 mg cm$^{-2}$. The samples were then thermally treated in a mixture of H$_2$ (5 sccm) and Ar (100 sccm) at 750°C for 1h. For comparison, the Pt/C catalysts were fabricated by a similar procedure, while 20 wt% Pt/C powder (15 mg) was dispersed in a mixture of isopropanol (4.2 mL), deionized water (1.5 mL), and Nafion solution (0.3 mL) to form the dispersion followed by sonication for 10 min and the dispersion was dropped into the Cu foam. All electrochemical experiments were conducted in standard three-electrode electrolyser in 1M KOH electrolyte, with the Hg/HgO electrode and Ni foam as the reference electrode and counter electrode. The scan rate was 1 mV s$^{-1}$ and an 85% *iR* correction was taken.

**Material characterization.** The samples were characterized by SEM (Hitachi SU8010, 5 kV), TEM (FEI Tecnai G2 F30, 300 kV and FEI Titan Themis G2, 60 kV), AFM (Oxford Instruments, Cyper ES, tapping mode), XPS (Thermo Fisher Scientific, ESCALAB Xi, monochromatic Al Kα X-rays, 1484.6 eV), XRD (Bruker D8 Advance, with monochromatic Cu Kα radiation λ = 0.15418 nm), Raman spectroscopy (Horiba, LabRAM ER, 532 nm laser excitation), UV-vis-NIR absorption spectrometer (SHIMADZU, UV-2600), DLS and zeta potential measurements (Malvern Zetasizer Nano-ZS90), TGA (Mettler Toledo, TGA2 Star System), and digital viscometer (Shanghai Lichen Bangxi Instrument Technology Co. Ltd., NDJ-8S). The thermal conductivity of the BNNS/CMC composite film was calculated from the thermal diffusion coefficient, density, and heat capacity. The thermal diffusion coefficient was measured by a laser flash apparatus (NETZSCH, LFA447). The density was measured by a density meter (Mettler Toledo, MAY-ME104), and the heat capacity was measured by a differential scanning calorimeter (Mettler Toledo, DSC 3). Mechanical tests of the BNNS/CMC composite film were carried out on a static testing instrument with two clamps (Instron 5943). The dielectric properties of the BNNS/CMC composite film were measured using wide-band dielectric spectroscopy (NOVOCONTROL, Concept 40).

**Data availability**

The data that support the figures within this paper and other findings of this study are available from the corresponding author upon reasonable request.

**Acknowledgments**

This work was financially supported by the National Natural Science Foundation of China (Nos. 51920105002, 51991340, and 51991343), China Postdoctoral Science Foundation (No.2020M680540), Guangdong Innovative and Entrepreneurial Research Team Program (No. 2017ZT07C341), the Bureau of Industry and Information Technology of Shenzhen for the "2017 Graphene Manufacturing Innovation Center Project" (No. 201901171523), and the Shenzhen Basic Research Project (Nos. JCYJ20190809180605522, JCYJ20200109144620815, and JCYJ20200109144616617). This work is also assisted by SUSTech Core Research Facilities, especially technical support from Pico-Centre that receives support from Presidential fund and Development and Reform Commission of Shenzhen Municipality. We would like to thank Mingqiang Liu for the synthesis of the $Bi_2O_2Se$ crystal, and Zhiyuan Zhang and Ziyang Huang for taking photos of the 10 L BNNS and $MoS_2$ dispersions

# References

[1] S. Witomska et al., Adv. Funct. Mater. 29 (22) (2019) 1901126.
[2] X. Cai et al., Chem. Soc. Rev. 47 (16) (2018) 6224.
[3] L. Yang et al., Nano Research 14 (6) (2021) 1583.
[4] R. Rizvi et al., Adv. Mater. 30 (30) (2018) 1800200.
[5] D. Shi et al., Small 16 (13) (2020) 1906734.
[6] N. Wang et al., Mater. Today 27 (2019) 33.
[7] C. X. Hu et al., Nanoscale 13 (2) (2021) 460.
[8] R. Tao et al., Adv. Funct. Mater. 31 (6) (2020) 2007630.
[9] S. Biccai et al., 2D Mater. 6 (1) (2018) 015008.
[10] A. G. Kelly et al., 2D Mater. 6 (4) (2019) 045036.
[11] J. N. Coleman et al., Science 331(6017) (2011) 568.
[12] K. R. Paton et al., Nat. Mater. 13 (6) (2014) 624.
[13] S. Chen et al., Adv. Mater. 31 (10) (2019) 1804810.
[14] D. Lee et al., Nano Lett. 15 (2) (2015) 1238.
[15] Z. Lin et al., Nature 562 (7726) (2018) 254.
[16] S. Yang et al., Angew. Chem. Int. Ed. 57 (17) (2018) 4677.
[17] Z. Chi et al., Natl. Sci. Rev. 7 (2) (2020) 9.
[18] K. S. Novoselov et al., Science, 306 (5696) (2004) 666.
[19] Y. Liu et al., Adv. Mater. 29 (44) (2017) 1703028.
[20] W. Zeng et al., Adv. Energy Mater. 8 (11) (2018) 1702314.
[21] C. Huang et al., Angew. Chem. Int. Ed. 58 (23) (2019) 7636.
[22] J. Peng et al., Matter 2 (1) (2020) 220.
[23] H. Liu et al., ACS Appl. Mater. Interfaces 11 (43) (2019) 40613.
[24] M. Qiu et al., Proc. Natl. Acad. Sci. U S A 115 (3) (2018) 501.
[25] J. Chen et al., ACS Nano 13 (1) (2019) 337.
[26] J. H. Jung et al., Nano Lett. 18 (5) (2018) 2759.
[27] G. Kresse et al., Phys. Rev. B, 54 (16) (1996) 11169.
[28] J. P. Perdew et al., Phys. Rev. Lett. 77 (18) (1996) 3865.
[29] G. Kresse et al., Phys. Rev. B 59 (3) (1999) 1758.
[30] S. Grimme et al., J. Chem. Phys. 132 (15) (2010) 154104.
[31] W. Lei et al., Nat. Commun. 6 (2015) 8849.
[32] Z. Zeng et al., Angew. Chem. Int. Ed. 51 (36) (2012) 9052.
[33] X. Hu et al., Adv. Mater. 29 (5) (2017) 1604031.
[34] X. F. Pan et al., Nat. Commun. 9 (1) (2018) 2974.
[35] H. Xie et al., Small 16 (1) (2020) 1905208.
[36] X. Zhang et al., Chem. Soc. Rev. 44 (9) (2015) 2757.
[37] C. Du et al., ACS Appl. Mater. Interfaces 10 (40) (2018) 34674.
[38] Y. Zhuang et al., ACS Nano 14 (9) (2020) 11733.
[39] V. Guerra et al., Prog. Mater. Sci. 100 (2019) 170.
[40] A. Akbari et al., Nat. Commun. 7 (2016) 10891.

[41] J. Han et al., Adv. Funct. Mater. 29 (13) (2019) 1900412.

[42] Y. Zhang et al., Polymer 143 (2018) 1.

[43] X. Zhang et al., ACS Appl. Mater. Interfaces 9 (27) (2017) 22977.

[44] T. Morishita et al., ACS Appl. Mater. Interfaces 8 (40) (2016) 27064.

[45] X. Zeng et al., Nanoscale 7 (15) (2015) 6774.

[46] Z. Kuang et al., Small 11 (14) (2015) 1655.

[47] H. Shen et al., ACS Appl. Mater. Interfaces 7 (10) (2015) 5701.

[48] Y. Luo et al., Nat. Commun. 10 (1) (2019) 269.

[49] C. Zhang et al., Nat. Commun. 11 (1) (2020) 3724.

[50] Y. Rao et al., ACS Appl. Mater. Interfaces. 12 (33) (2020) 37092.

[51] C. Jian et al., Appl. Catal. B: Environ. 266 (2020) 118649.

[52] L. Wang et al., ACS Appl. Mater. Interfaces 11 (31) (2019) 27743.

[53] L. Yu et al., Nano Energy 53 (2018) 492.

[54] X. Zou et al., Chem 4 (5) (2018) 1139.

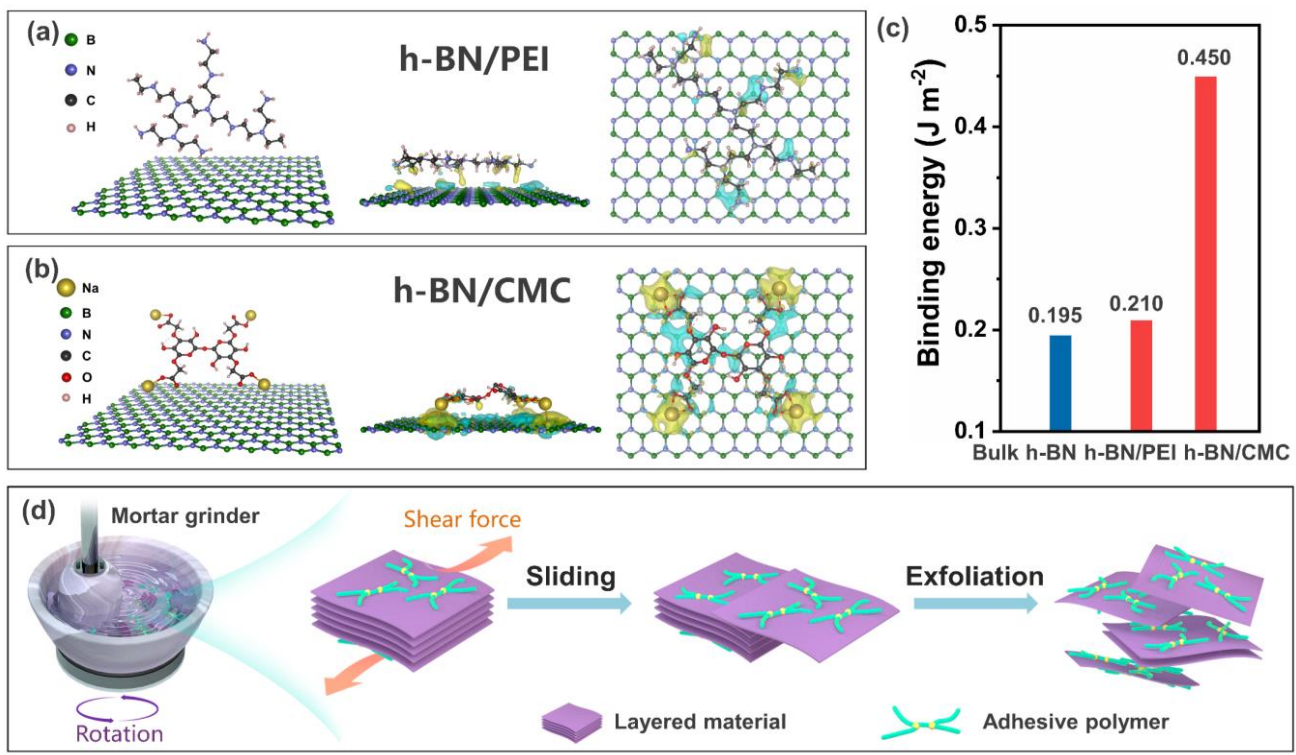

**Figure 1. Mechanism and schematic of the GAGE method.** Structure, side and top views of charge density difference of (a) h-BN/PEI, and (b) h-BN/CMC. (c) Comparison between the binding energies of h-BN/polymer and the exfoliation energy of bulk h-BN. (d) Schematic of the GAGE process for the production of 2D materials.

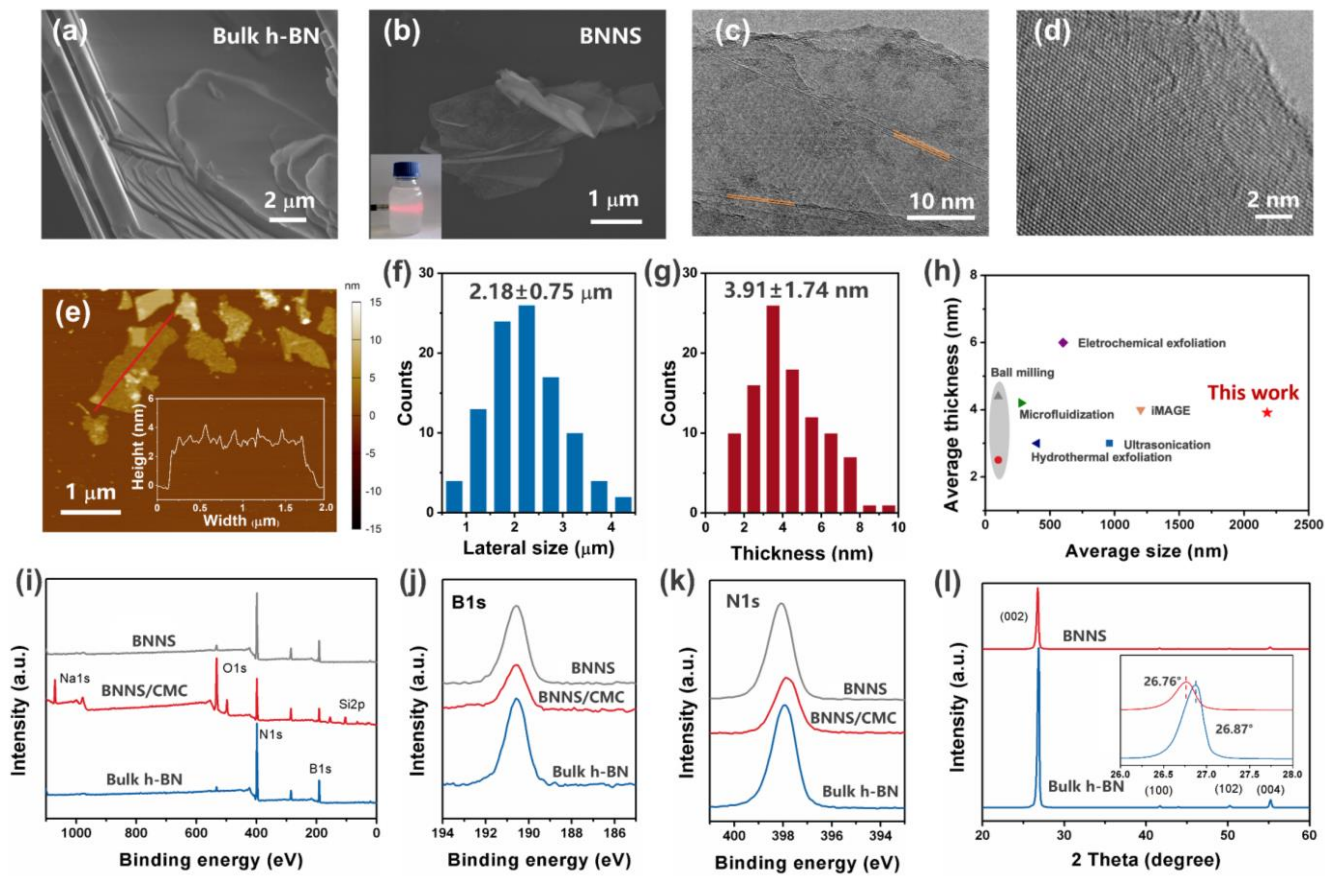

**Figure 2. Characterization of the GAGE produced BNNSs.** SEM images showing (a) bulk BN powders, (b) BNNSs after the GAGE process, inset is a photo of a BNNS dispersion showing the Tyndall effect. (c, d) HRTEM images of BNNSs. (e) AFM image of the BNNSs and the corresponding height profile (inset). (f) Lateral size, (g) thickness histogram of BNNS. (h) The average size and thickness of BNNSs reported in the literature and this work. (i) Survey XPS spectra, and high-resolution XPS spectra of (j) the B 1s peaks and (k) the N 1s peaks from bulk h-BN, BNNS-CMC, and BNNS. (l) XRD characterization of the bulk h-BN and BNNS.

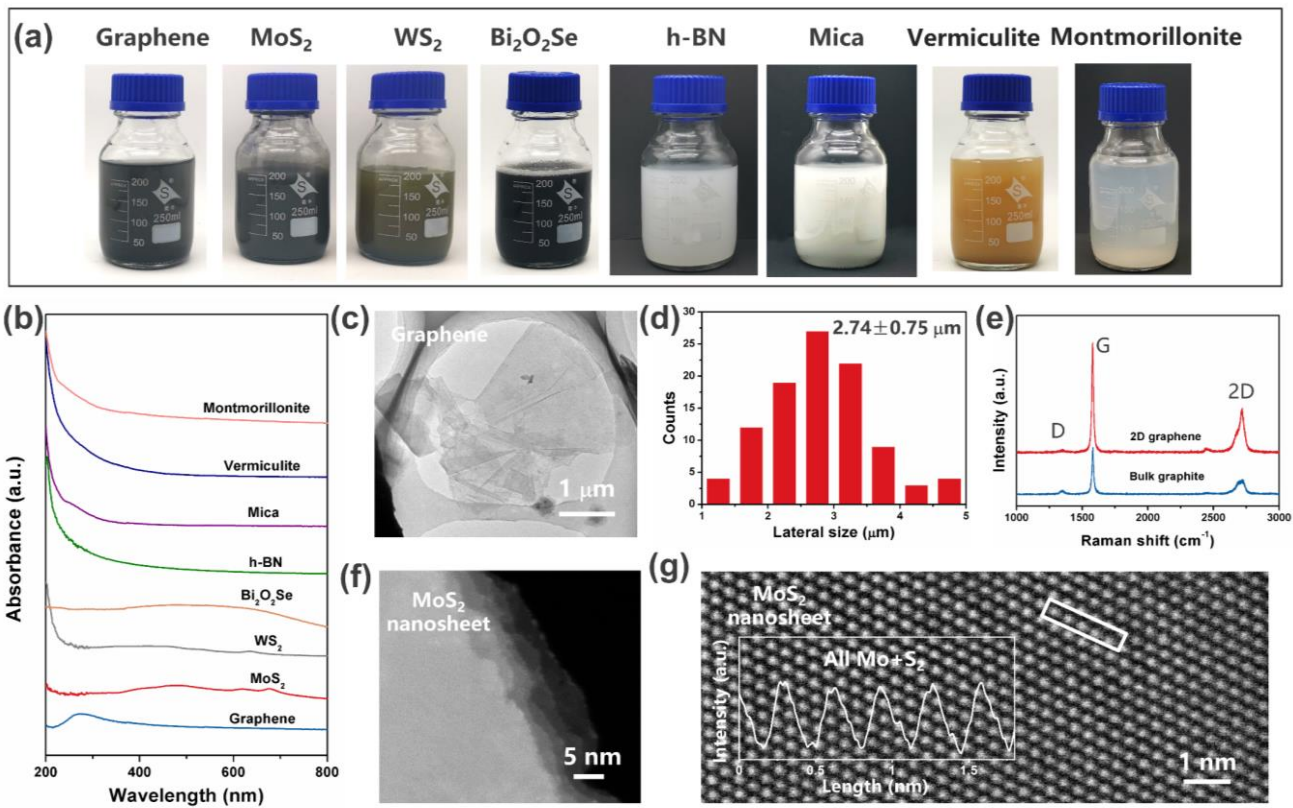

**Figure 3. The universality of the GAGE method in producing different 2D materials.** (a) Photos and (b) UV-vis‑NIR absorption spectra of the GAGE produced 2D sheet dispersions. (c) TEM image and (d) histogram in the lateral size of the GAGE exfoliated graphene. (e) Raman spectra of graphene sheets and graphite powders. HAADF-STEM images of $MoS_2$ nanosheets showing (f) clear separated edges, and (g) high crystalline quality. Inset is the intensity line profile of the selected atom columns in the white rectangle.

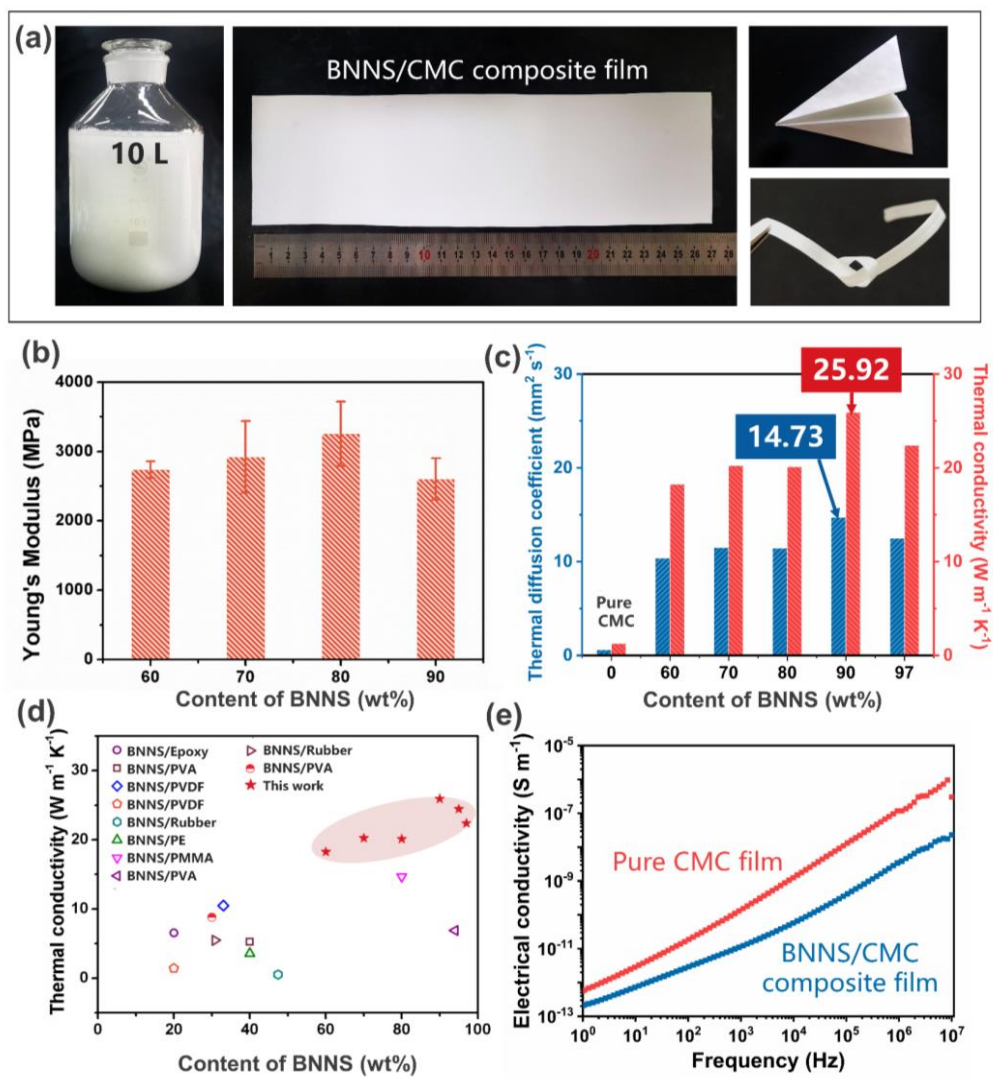

**Figure 4. Applications of the GAGE produced BNNS in insulating thermal conduction.** (a) Photos of a 10L BNNS/CMC dispersion and a BNNS/CMC composite film with an area of ~ 200 cm$^2$ and good flexibility at a BNNS loading of 60 wt%. (b) Young's Modulus of the BNNS/CMC composite film with different BNNS loadings ranging from 60 to 90 wt%. (c) Thermal conductivity and thermal diffusion coefficient of the pure CMC film and BNNS/CMC composite film with different BNNS loadings ranging from 60 to 97 wt%. (d) Thermal conductivity of representative BNNS/polymer composites and this work. (e) The frequency-dependent electrical conductivity of a pure CMC film and the BNNS/CMC composite film.

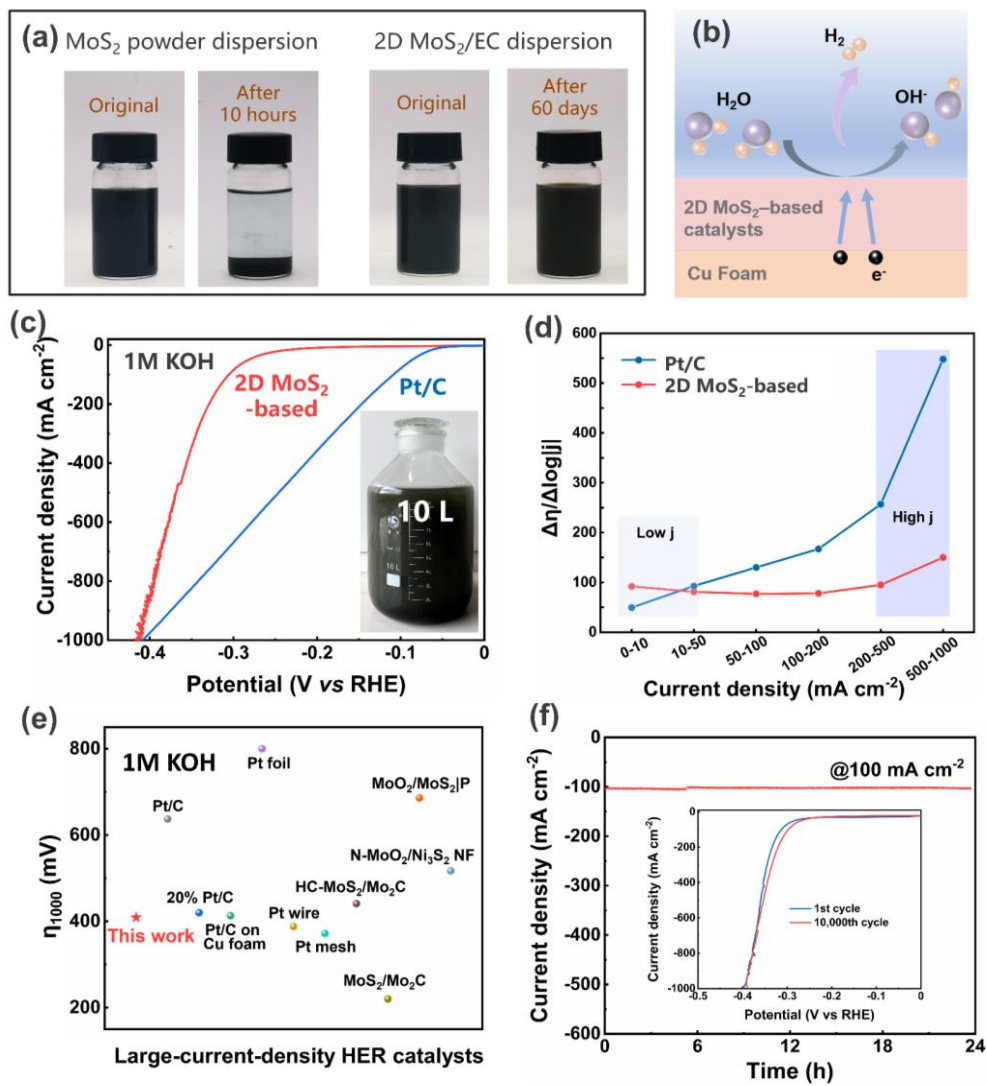

**Figure 5. Applications of the GAGE produced 2D MoS₂ in large-current-density HER.** (a) Photos of MoS$_2$ powder dispersion showing obvious precipitates after 10 hours and 2D MoS$_2$/EC dispersion showing its retention of dispersibility after 60 days. (b) Schematic of the 2D MoS$_2$-based HER. (c) Polarization curves with *iR* compensation and (d) $\Delta\eta/\Delta\log|j|$ ratio of 2D MoS$_2$-based catalysts and Pt/C catalysts in 1M KOH electrolyte, inset is a photo of the 10 L 2D MoS$_2$ dispersion. (e) Overpotential at a large current density of 1000 mA cm$^{-2}$ of reported Pt-based catalysts and MoS$_2$-based composite catalysts, as well as this work. (f) Chronoamperometric responses (i–t) recorded on a 2D MoS$_2$-based catalyst for 24 h at 100 mA cm$^{-2}$. Inset shows the polarization curves of the 2D MoS$_2$-based catalyst during the initial scan and after 10,000 scans.

23